\documentclass[letterpaper, 10 pt, conference]{ieeeconf}
\IEEEoverridecommandlockouts    
\usepackage{graphicx}
\usepackage{times}              
\usepackage{amsmath}            
\usepackage{amssymb}            
\usepackage{gensymb}
\usepackage{wrapfig}
\usepackage{xcolor}
\usepackage{subcaption}
\usepackage{algorithm}
\usepackage{algpseudocode}
\usepackage{array}              
\newcolumntype{P}[1]{>{\centering\arraybackslash}p{#1}}

\title{\LARGE \bf Use of Quadcopter Wakes to Supplement Strawberry Pollination}
\author{Sadie Cutler$^{1}$, Benjamin Defay$^{2}$, Scott McArt$^{3}$, and Kirstin Petersen$^{2}$
\thanks{Cornell Institute for Digital Agriculture (CIDA)}\\
\thanks{$^{1}$Sibley School of Mechanical and Aerospace Engineering, Cornell University, Ithaca, NY 14853, USA
        {\tt\small sc3236@cornell.edu}}%
\thanks{$^{2}$School of Electrical and Computer Engineering, Cornell University, Ithaca, NY 14853, USA
        {\tt\small kirstin@cornell.edu}}%
\thanks{$^{3}$College of Agriculture and Life Science, Cornell University, Ithaca, NY 14853, USA
        {\tt\small shm33@cornell.edu}}%
}

\begin{document}
\maketitle
\thispagestyle{empty}
\pagestyle{empty}

\begin{abstract}
Pollinators are critical to the world's ecosystems and food supply, yet recent studies have found pollination shortfalls in several crops, including strawberry. This is troubling because wild and managed pollinators are currently experiencing declines. One possibility is to try and provide supplemental pollination solutions. These solutions should be affordable and simple for farmers to implement if their use is to be widespread; quadcopters are a great example, already used for monitoring on many farms. This paper investigates a new method for artificial pollination based on wind pollination that bears further investigation. After determining the height where the lateral flow is maximized, we performed field experiments with a quadcopter assisting natural pollinators. Although our results in the field were inconclusive, lab studies show that the idea shows promise and could be adapted for better field results.
\end{abstract}

\section{Introduction}
\label{sec:intro}

Pollinators are crucial to the world's ecosystem and food supply; 87 of the 115 of the primary crops grown for consumption worldwide require the work of natural pollinators to develop~\cite{klein2007importance}. There are ~20,000 species of bees worldwide and 4,337 in North America; roughly 1/3 of the food we eat is pollinated by these bees~\cite{beeconservancy_whybees}. Over the past several decades, however, an increase in pesticides, imported species and disease, climate change, pollution, and shrinking habitats have led to a decline of more than 50\% in North American bee species (including managed bees), with 24\% being in danger of extinction~\cite{brunet2024main, nath2023insect}. To combat this, an established practice is for farmers to introduce managed hives to their fields, but studies show that managed hives (predominantly honey bees) can harbor pathogens that are harmful to wild bees, in addition to the competition they pose to natural resources~\cite{brunet2024main}. The decline of natural, wild pollinators therefore, is a problem. While natural pollinators will hopefully always play the primary role in pollination of crops and other plants, a robotic solution could play a key role in ensuring food supply for a growing population while research to save and rebuild wild bee populations is conducted in tandem. 

This work proposes implementing a robotic solution that can work alongside bees without endangering them. We propose using the lateral flow generated by a quadcopter's downdraft to aid in cross-pollination of strawberry plants. We used a small recreational quadcopter (DJI Phantom 2) to test the efficacy of the idea in lab experiments and in a large field study, using weight and computer vision analysis to determine pollination rates of each strawberry after harvest. Although our initial field results are not as pronounced as we had hoped, our preliminary lab experiments show that the method has promise and the prevalence of recreational quadcopters as well as quadcopters integrated in other aspects of farming, make it a low-cost and time-efficient addition to farming.

Our work was conducted using strawberry plants, which contribute 8.5 million dollars to New York state's GDP (where this research was conducted)~\cite{Cornell_Strawberries} and roughly \$2 billion nationwide~\cite{Yeh2023}. Strawberries grow low to the ground, flower for several weeks, are typically highly impacted by pollination. In addition, their markers for pollination have been well-studied~\cite{klatt2014bee}. Although strawberries are primarily pollinated by natural pollinators, they are also wind- and self-pollinated which, together with their earlier mentioned traits,  makes them an ideal candidate for our method of supplemental artificial pollination.

This paper details the lab experiments that were conducted to help determine parameters for the field work, as well as a correlating field study that tested our hypothesis. 
Sec.~\ref{sec:lab} outlines the experimental design, and presents and discusses the results of the experiments conducted in the lab.
Sec.~\ref{sec:field} explains the methods used for field experiments, including the physical setup, as well as the code used to analyze the data; it also includes the experimental results and their discussion.
Sec.~\ref{sec:conclusion} reviews the implications of the data and how we can use it to re-design future experiments.
This work is a first step towards improved execution of robotic solutions for assisting natural pollinators in crop pollination.

    \begin{figure}[b]
        \centering
        \includegraphics[width=\columnwidth]{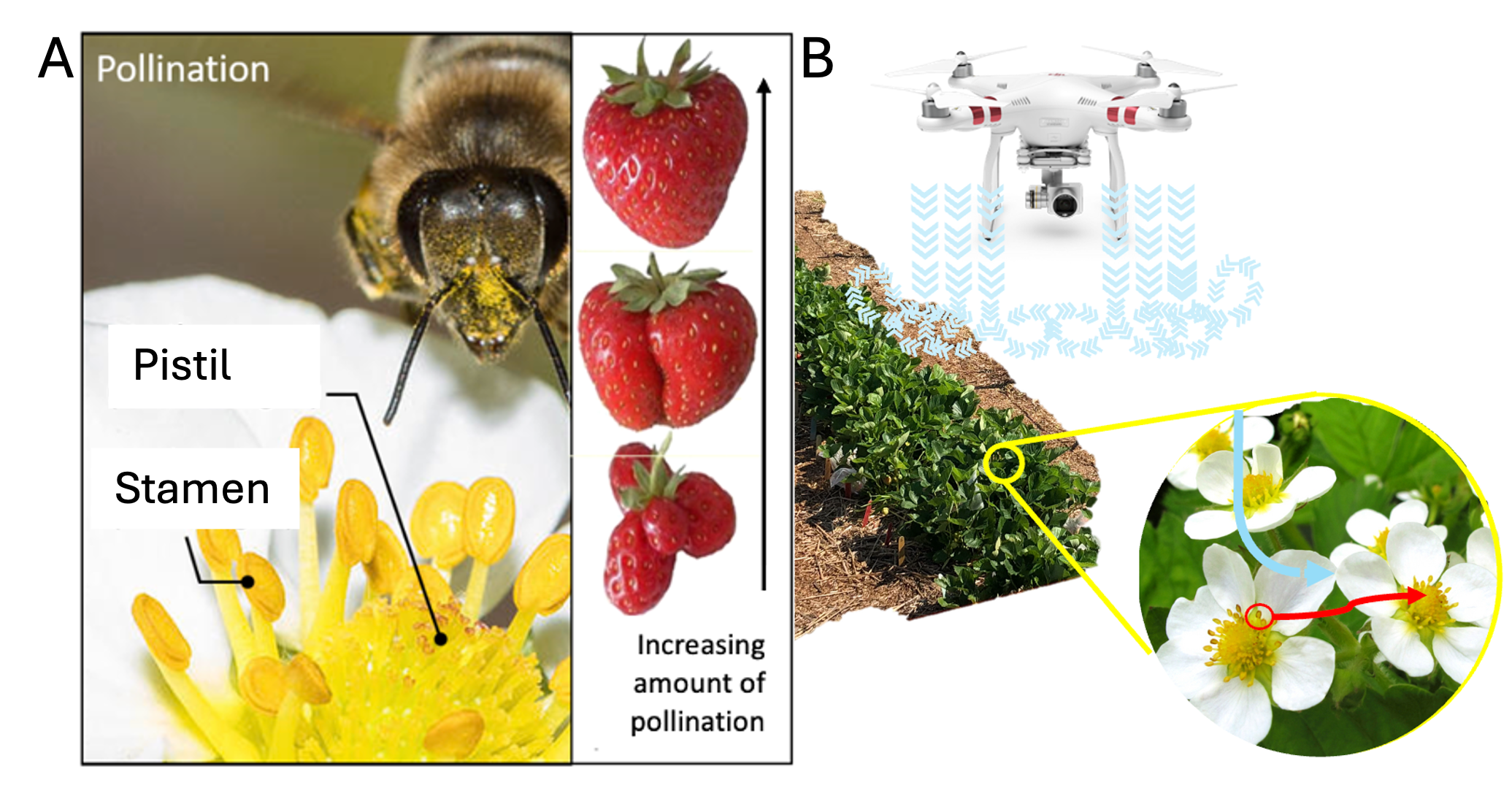}
        \caption {\small{A) pollen from the anther (part of the stamen) is transferred to the stigma (part of the pistil), which then develop flesh around them and become achenes. The more achenes are pollinated, the heavier and more shapely the fruit will be; B) example of how the quadcopter's downdraft that potentially could be a mechanism to transfer the pollen.}}
        \label{fig:pollination_wake}
    \end{figure}

\section{Related Work}
\label{sec:relatedworks}


Perhaps the simplest solution to an under-pollinated crop is to hand-pollinate. Hand-pollination has been in practice for hundreds of years and is still used in some parts of the world today. Although this solution does not require expensive tools, it does require a tremendous amount of man-power, which is both costly and in high demand.~\cite{partap2012human, broussard2023artificial,USDA2025FarmLabor} Although hand-pollinating flowers is uncommon, some growers have workers use electrostatic, liquid-, or air-based sprayers to spray pre-harvested pollen in the hope of increasing yields~\cite{broussard2023artificial}. Research on the efficacy of these approaches and whether they can be automated has yielded a number of robotic alternatives.

To date, automated robotic solutions to under-pollination include air-jets, sprays, vibrations, ultrasonic waves, and linear actuators~\cite{singh2024comprehensive}. However, the diverse anatomy of plants, pruning patterns, and planting spacing limits the universality of solutions even where cost is not prohibitive. The approaches form two main categories: those that pollinate flowers individually and those that use a less precise 'spray and pray' approach.

Pollinating individual flowers is an approach that can be combined with either spraying pre-harvested pollen or directly transferring the pollen between flowers~\cite{broussard2023artificial}. A robot arm is typically mounted on a driving platform; computer vision and machine learning allow the robot arm to identify flowers; and trajectory planning is executed so the end effector can either spray the pollen onto the flower~\cite{gao2023novel} or execute a brushing motion~\cite{li2022design,ohi2018design} to move the pollen from the stamen to the pistils. While finding each individual flower is optimal for pollination, it is a very slow process, which is prone to occlusions unless robot-friends pruning practices are implemented. In addition, the end effector may risk damaging any part of the flower while interacting with it.
        
In contrast, some robotic pollination efforts have found success without the need to individually identify flowers.
One study mounted a path of animal hair coated in ionic liquid gel on a quadcopter which mimicked how pollinators pick up and deposit pollen when they come in contact with flowers~\cite{chechetka2017materially}. 
Generally the approaches that do not locate individual powers take advantage of pre-harvested pollen though~\cite{broussard2023artificial}. 
For example, the pre-harvested pollen can be combined with soap in a liquid mixture to create pollen-infused soap bubbles. By mounting this bubble machine on a quadcopter, the quadcopter's downdraft sends the pollen bubbles towards flowers below. The bubbles pop on contact with the pistils, delivering the pollen without injuring the flower.~\cite{yang2020soap}. This works well for flowers like lilies or pears, which need only a few pistils to be pollinated to produce fruit~\cite{beeaware_apples_pears} and whose pistils protrude from the plant, but would be unlikely to translate to something like strawberries, where 100-400 pistils need to be pollinated for the fruit to develop properly or soybeans whose pistils are hidden inside the flower~\cite{manitoba_agriculture_strawberry_plant, ricciuti2024beesounce}. 
Quadcopter delivery can also be used when the pollen is combined with a heavier powder; the powder can be released gradually into the wake of a quadcopter so that the quadcopter's downdraft spreads the pollen onto the waiting flowers below; this practice has been adopted commercially by companies like DropCopter$^{\textregistered}$~\cite{mazinani2021design,fine2019apparatus}. While the approach seems promising, it has not been studied for effectiveness.

The use of contactless pollination is advantageous as it is less time-consuming, has a lower risk of flower damage, and has less dependence on the presence of occlusions. In addition, the use of quadcopters avoids complications related to rough terrain traversal by ground-based robots.
Our approach builds on the work done with quadcopter downdrafts and the idea that a solution that works for farmers should be affordable and easy to implement.
Inspired by the quadcopter modeling of downdrafts done by~\cite{lee2020rotor, feng2018measurement}, we hypothesize that the wind interactions that occur at the ground could result in pollen transfer between flowers without the need for pollen to be injected into the downdraft.

\section{Lab Experiments}
\label{sec:lab}

\subsection{Methods}
\label{sec:lab_methods}

    \begin{figure}[b!]
        \centering
        \includegraphics[width=.8\columnwidth]{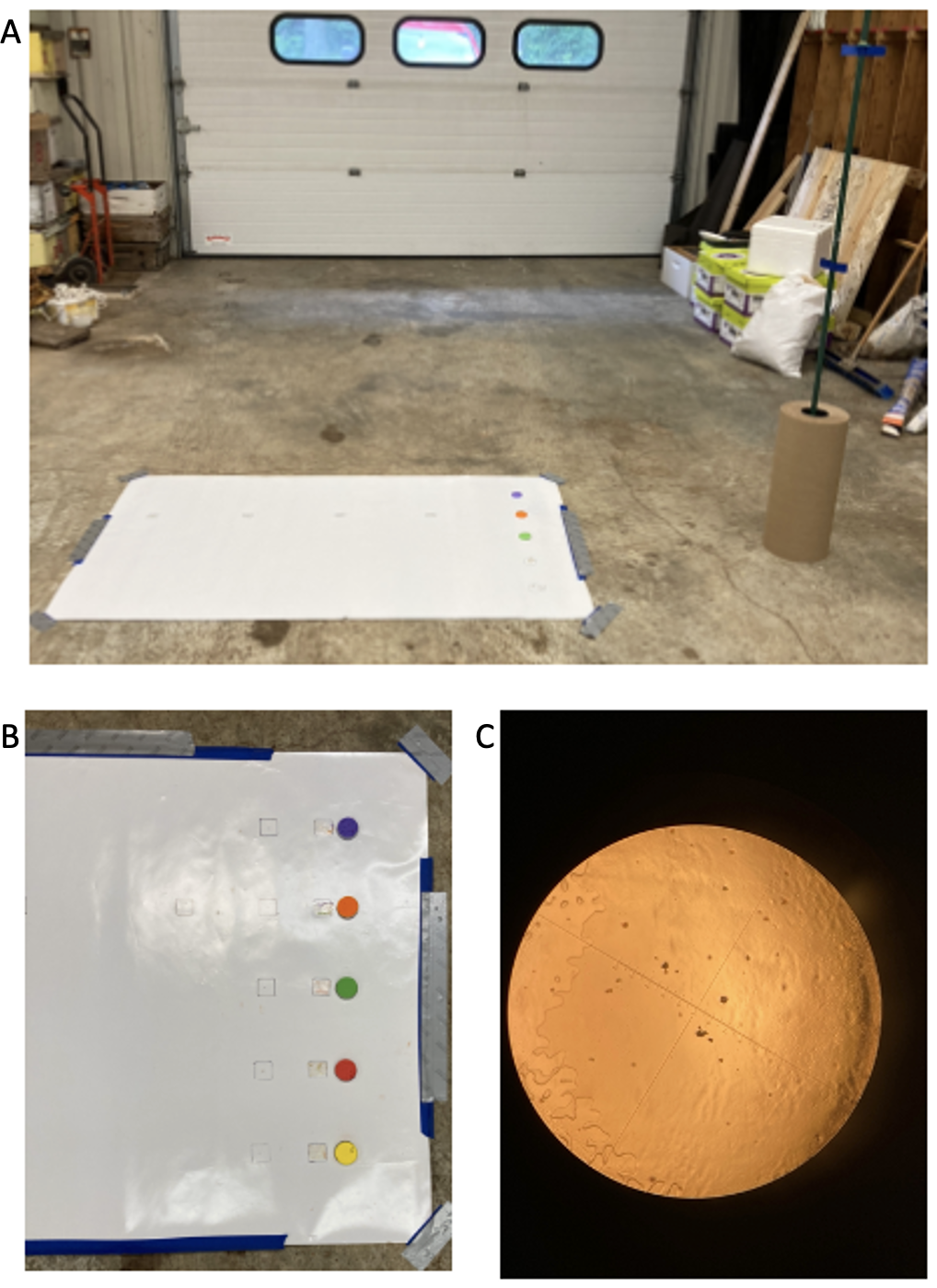}
        \caption {\small{A) Setup to investigate the effects of height on pollen transfer with tape on the pole indicating the experimental height; B) Setup to investigate the effects of flight pattern on pollen transfer (the height marker was also used but remained at the same height); C) an example of what the simulated pollen looked like on the glass slide.}}
        \label{fig:dycesetup}
    \end{figure}
    
To devise a rigorous way to determine the optimal height of the quadcopter for maximizing lateral pollen transfer, we conducted experiments in the lab substituting Holi powder (particle size: 5-30$\mu m$) for strawberry pollen (particle size: 15-40$\mu m$) because of its availability and similar scale~\cite{maas1977pollen,bossmann2016holi}. The quadcopter used in all experiments described in this paper was a DJI Phantom II. 
    
For the first experiment, we placed five 100 mm x 15 mm Petri dishes of Holi powder in a row (spaced 150 mm apart) and then placed four 25 mm square glass slides coated with double-sided tape in 300 mm increments in an orthogonal row adjacent to the middle Petri dish as seen in \ref{fig:dycesetup}A. We then flew the quadcopter back and forth directly over the row of Petri dishes three times per trial, conducting three trials each at .6 m, .9 m, 1.2 m, and 1.5 m (height increments were chosen in feet and later converted).
    
For the second experiment, we again placed five 100 mm x 15 mm Petri dishes of Holi powder in a row spaced 150 mm apart and then placed two rows of sticky glass slides parallel to the Petri dishes with the first row 25 mm away, and the second row 75 mm away from the Petri dishes, with each slide in line with a Petri dish as seen in \ref{fig:dycesetup}B. In this experiment, we maintained a height of 1.2 m and varied the flight patterns: straight (i.e. flying quickly over the row), hovering (i.e. flying extremely slowly over the row), and zig-zagging (i.e. switching quickly between forward and backward motion over the row in a rocking motion). We performed three trials for each flight pattern.

    \begin{figure}[b!]
        \centering
        \includegraphics[width=.8\columnwidth]{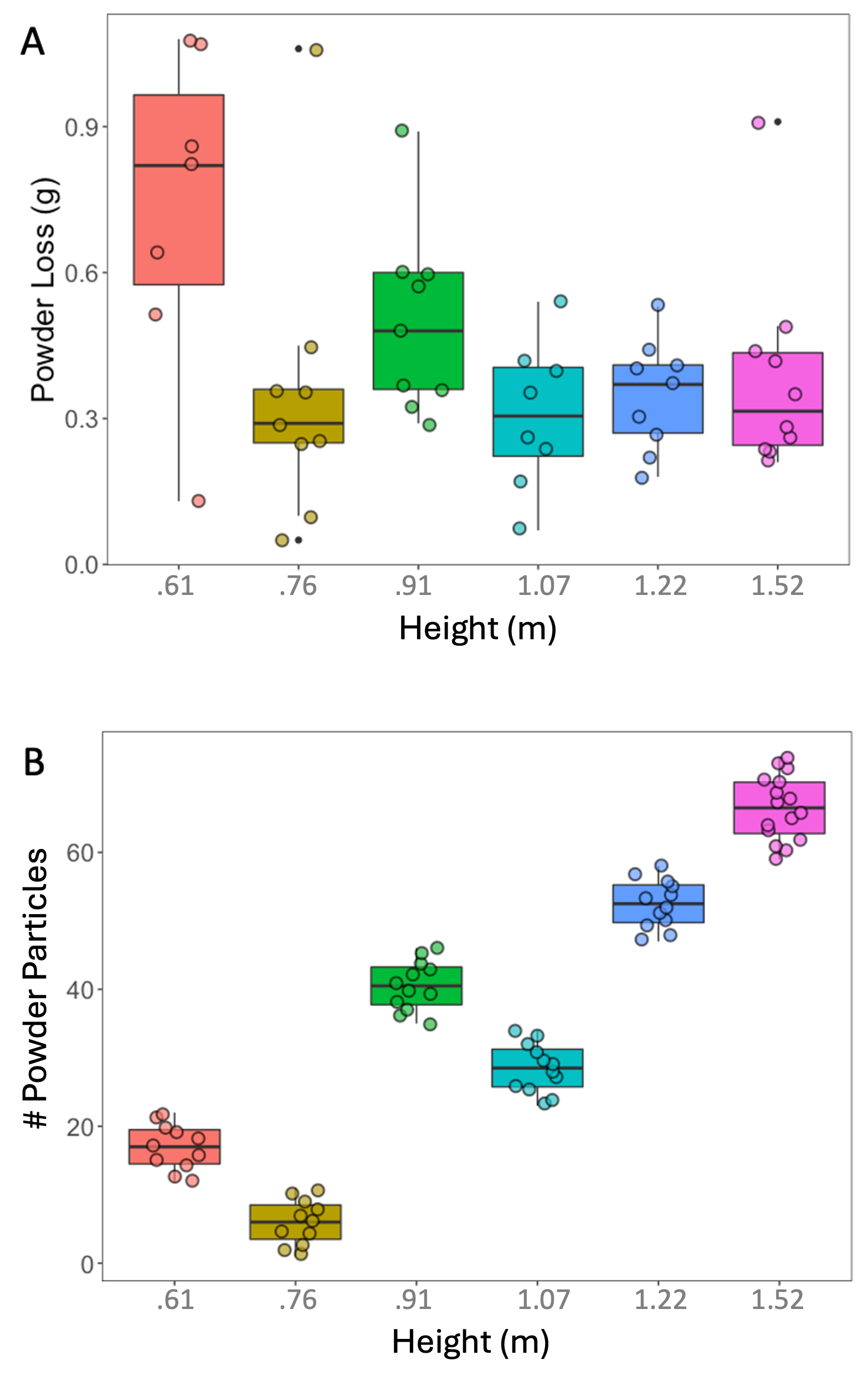}
        \caption {\small{Box and whisker plots with data overlaid for the first experiment investigating the effects of height on pollen transfer A) Holi powder loss mass measurements and B) Holi powder transfer as number of particles}}
        \label{fig:dycesetupH}
    \end{figure}
    
For both sets of experiments, we measured the mass of the Petri dishes filled with Holi powder before and after each trial and recorded the difference. We also took images of the center of each slide under a microscope, noting its placement in the experiment, and manually counted the Holi particles that had adhered to the sticky slide in the same magnified area.
    
To analyze the data, we created linear mixed models of the data for: 1) powder loss (difference in mass of Holi powder before and after each trial) and 2) powder transfer (accumulation of powder particles on the sticky slides). 
We used the treatment group as our fixed effect for all the linear mixed models.
For the height experiment, the random effect for model (1) was Petri dish position, and for model (2) was the trial number. 
For the flight pattern experiment, the random effects for model (1) were Petri dish position and trial number, and for model (2) were slide position and trial number.
We then did pairwise comparisons on the different treatment groups using the Kenward-Roger method and 95\% confidence intervals and calculated the power of the experiment.

\subsection{Results}
\label{sec:lab_results}

    \begin{figure}[b!]
        \centering
        \includegraphics[width=.7\columnwidth]{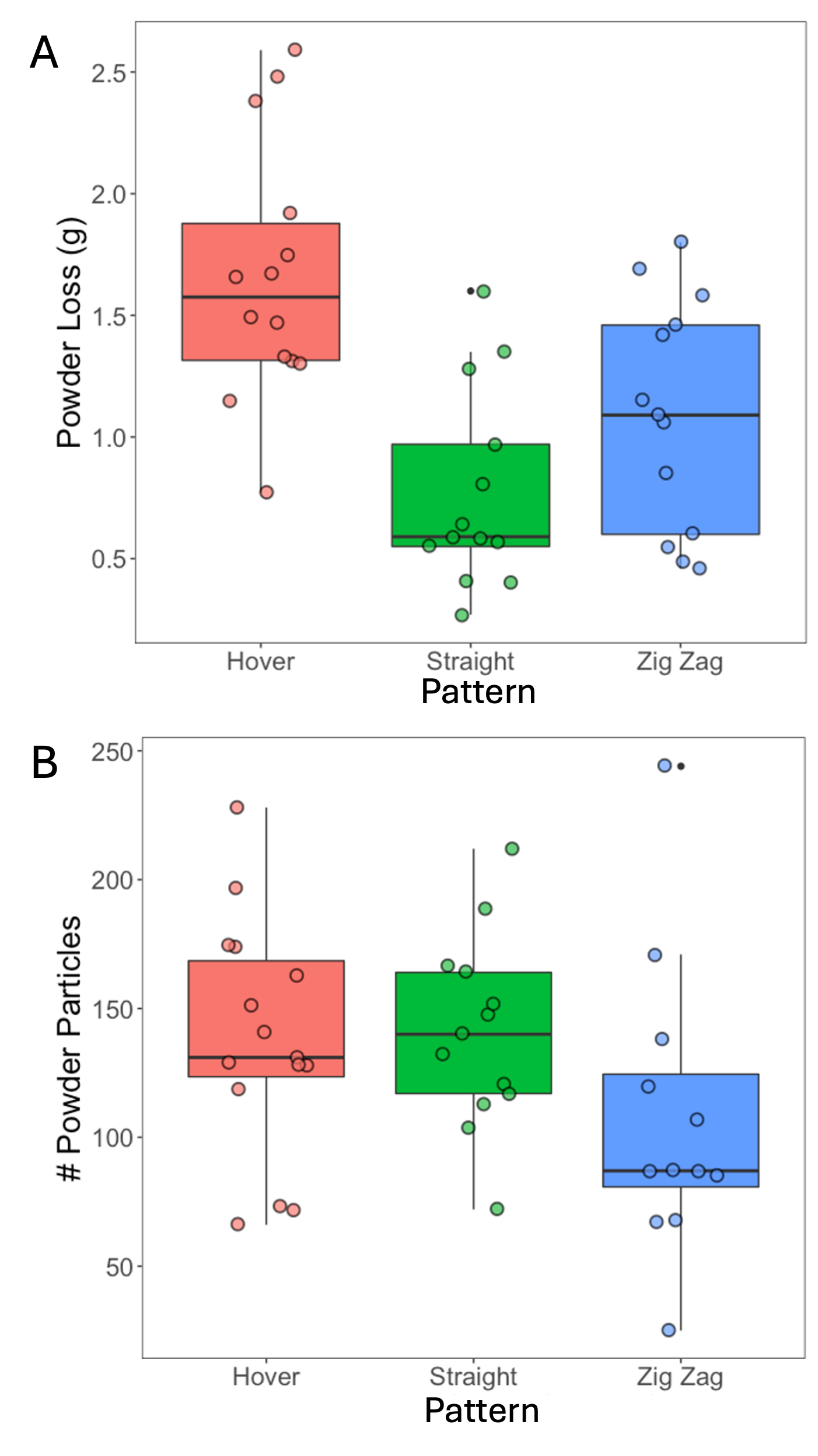}
        \caption {\small{Box and whisker plots with data overlaid for the experiment investigating the effects of flight pattern on pollen transfer. A) simulated pollen loss mass measurements and B) simulated pollen transfer as number of particles.}}
        \label{fig:dycesetupP}
    \end{figure}

For the experiment investigating the quadcopter wake's effect on simulated pollen movement at different heights, the power of the experiment was 1. 
The pairwise comparisons for the linear mixed model investigating powder loss at different heights (see Fig.\ref{fig:dycesetupH}A) show statistical significance (p-values less than 0.05) in the (.61 m, .76 m), (.61 m, 1.07 m), (.61 m, 1.22 m), and (.61  m, 1.52 m) pairs. 
The pairwise comparisons for powder transfer at different heights (see Fig.\ref{fig:dycesetupH}B), show statistical significance between every height pair.
        
For the experiment investigating the quadcopter wake's effect on simulated pollen movement using different flight patterns, the power of the experiment was 0.403. 
The pairwise comparisons for the linear-mixed model investigating powder loss for different flight patterns show statistical significance between all pattern pairs. 
The pairwise comparisons for powder transfer with different flight patterns show statistical significance between none of the pattern pairs.

\subsection{Discussion}
\label{sec:lab_discussion}

The lab experiment -- designed to choose a height and flight pattern for the quadcopter before conducting field work -- yielded promising results. 
The results from the height experiment indicate that 60 cm is the height at which the most simulated pollen would be removed from the strawberry flowers; as the quadcopter's height increases, there is less pollen displaced in a somewhat linear relationship. The height experiment also shows, however, that at greater heights more simulated pollen particles are transferred to the glass slides. Without modeling the fluid dynamics of the quadcopter's downdraft, one possible explanation of this seeming dichotomy is that at lower heights the quadcopter displaces more pollen, but it does not necessarily get carried laterally along the ground where the strawberry flowers would be; the quadcopter will have a larger, albeit weaker, wake relative to the ground at greater heights. So, for optimal pollen transfer, the quadcopter pilot would have to balance these two competing parameters. The particle size of the simulated particles and pollen is similar, but strawberry pollen is slightly denser than the simulated particles and therefore may behave less favorably and would certainly be less predictable in the presence of ambient wind.
The results from the flight pattern experiment show that the most pollen is displaced utilizing the hover method. This was our predicted outcome as the quadcopter is spending more time above each "plant". The levels of overall pollen transfer also appear to be highest with this flight pattern. Although we found no statistical significance in this experiment, these data led us to choose a hover pattern for the quadcopter flight and a height of 1.2 m valuing pollen transfer over pollen loss.

\section{Field Experiment}
\label{sec:field}

    \begin{table*}[h!]
        \centering
        \begin{tabular}{|l|l|l|l|l|l|l|l|l|l|l|l|}
        \hline
            &
            \begin{tabular}[c]{@{}l@{}}Mass\\ (g)\end{tabular} &
          \begin{tabular}[c]{@{}l@{}}Mass\\ $\sigma$ \end{tabular} &
          \begin{tabular}[c]{@{}l@{}}Area\\ ($in^2$)\end{tabular} &
          \begin{tabular}[c]{@{}l@{}}Area\\ $\sigma$ \end{tabular} &
          \begin{tabular}[c]{@{}l@{}}Sym.\\ (\%)\end{tabular} &
          \begin{tabular}[c]{@{}l@{}}Sym.\\ $\sigma$ \end{tabular} &
          \begin{tabular}[c]{@{}l@{}}Achene\\ Size \\ (px) \end{tabular} &
          \begin{tabular}[c]{@{}l@{}}Achene\\ Size\\ $\sigma$ \end{tabular} &
          \begin{tabular}[c]{@{}l@{}}Achene\\ Distance \\ (px) \end{tabular} &
          \begin{tabular}[c]{@{}l@{}}Achene\\ Distance\\ $\sigma$ \end{tabular} &
          \# Berries \\
        \hline
        Quadcopter+Bees & 11.84 & 6.44 & 30.96 & 9.65 & 4.12 & 2.6 & 30.68 & 6.83 & 163.52 & 24.21 & 142 \\
        \hline
        Quadcopter      & 10.06 & 5.93 & 27.23 & 9.12 & 4.40 & 2.7 & 26.78 & 7.29 & 153.28 & 25.37 & 154 \\
        \hline
        Bees       & 10.79 & 5.744 & 29.47 & 8.96 & 4.42 & 2.9 & 30.65 & 6.87 & 157.2 & 23.85 & 121 \\
        \hline
        Neither    &  9.21 & 6.08 & 26.767 & 9.99 & 4.81 & 3.3 & 25.69 & 7.24 & 153.22 & 24.70 & 114 \\
        \hline
        \end{tabular}
        \caption{\small{Breakdown of strawberries harvested in the four control groups and the average and standard deviation of the mass, projected 2D area, symmetry, achene size, and relative distance between achenes for the field experiment.}}
        \label{tab:2020Fdata}
    \end{table*}

\subsection{Methods}
\label{sec:field_methods}
    
Malwina, the strawberry plants used, is one of the later-blooming varieties~\cite{sonsteby2020extreme}; we started the experiment on May 28 2020, began harvesting on June 27, and finished harvesting on July 8.

\begin{figure}[b!]
        \centering
        \includegraphics[width=.8\columnwidth]{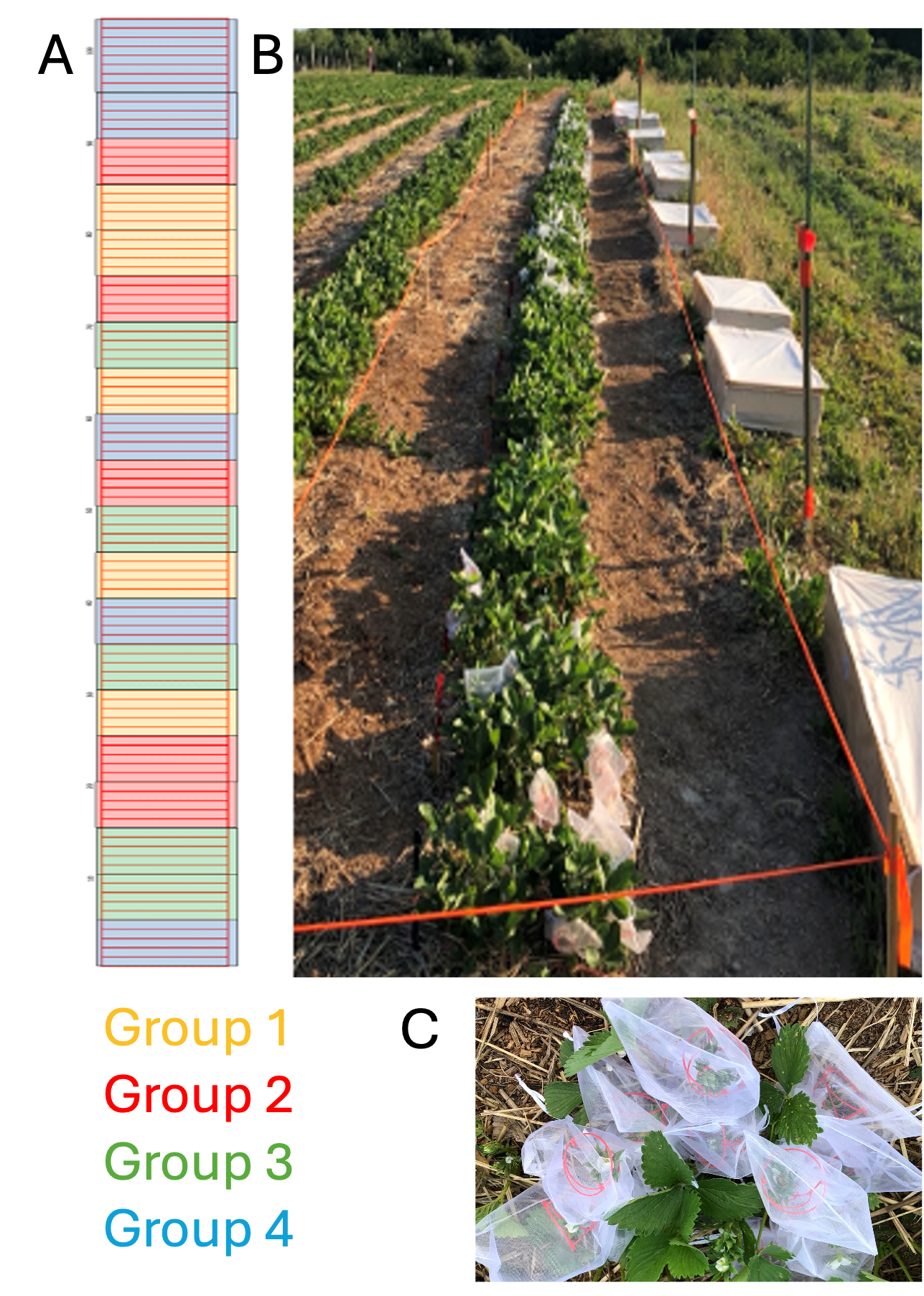}
        \caption {\small{A) Color-coded key showing which plants in the row of the field experiment were randomly assigned to which control groups; B) Example of the field setup and hardware used on any given day for the field experiment.}}
        \label{fig:fieldsetup}
    \end{figure}

Based on the data reference in Sec.~\ref{sec:lab_results}), for the field experiment, the quadcopter was flown slowly and steadily (hover pattern), at a height of ~1.2 m, directly over the row of strawberry plants over and back. Stakes were placed along the row every 3 m with visual markers applied at .9 m and 1.5 m to visually verify the quadcopter was in the range of acceptable heights since the quadcopter was operated manually (as seen in Fig.~\ref{fig:fieldsetup}).

The field experiment was conducted at Indian Creek Farms 42\degree28'12"N and 76\degree32'24"W. For consistency, all work for this experiment was carried out by a single researcher. We used one 15 m row of 100 Malwina strawberry plants, originally planted with 150 mm spacing; because the plants were mature and inter-grown, we sectioned with string every 150 mm to delineate sections and make it clear during harvest which section each berry belonged to. The 100 plants were blocked into groups of five and then assigned randomly to a control group as seen in Fig. \ref{fig:fieldsetup}. All the flowers used in this experiment were primary flowers, which can be determined simply by looking at the flower branching and will yield the largest berries.

We designated four control groups: 
\begin{itemize}
    \item Group 1: strawberries exposed to quadcopter pollination and wind pollination   
    \item Group 2: strawberries exposed to pollination by natural pollinators and wind pollination  
    \item Group 3: strawberries exposed to pollination by natural pollinators, quadcopter pollination, and wind pollination 
    \item Group 4: strawberries exposed only to wind pollination    
\end{itemize}

The plants in the control groups that were isolated from the quadcopter's wake were covered with long wooden boxes before flying the quadcopter; the boxes featured loose canvas on the short ends that rested between plants and topped with plastic sheeting which was elevated above the plants to prevent damaging the flowers. 
The plants in control groups that were isolated from natural pollinators had their flowers individually enclosed in a small mesh organza drawstring bag (13 x 18 cm) that allowed airflow but prevented natural pollinators (a standard method used by our Cornell bee collaborators). The mesh bags were installed with a spacer placed around the stem to hold the bag away from the flower.

Each of the four control groups contained 25 plants, and we tagged 200 primary flowers from each control group, resulting in 114-154 strawberries per control group at harvest (unharvested strawberries were lost to critters or disease).

Each morning upon arriving at the field, the researcher placed the boxes over the appropriate sections, flew the quadcopter twice over the row, and removed the boxes. The researcher also tagged and, if appropriate, bagged, any primary flowers that had bloomed since the previous day until the 200 flowers/control group target was reached. The quadcopter was flown daily until the last primary flower in the experiment had finished blooming.

Peak ripeness was defined as when all the flesh of the berry had turned from white to red. Upon reaching this state, berries were picked and carefully transported back to the lab, each labeled with the appropriate plant number.

    \begin{figure}[t!]
        \centering
        \includegraphics[width=\columnwidth]{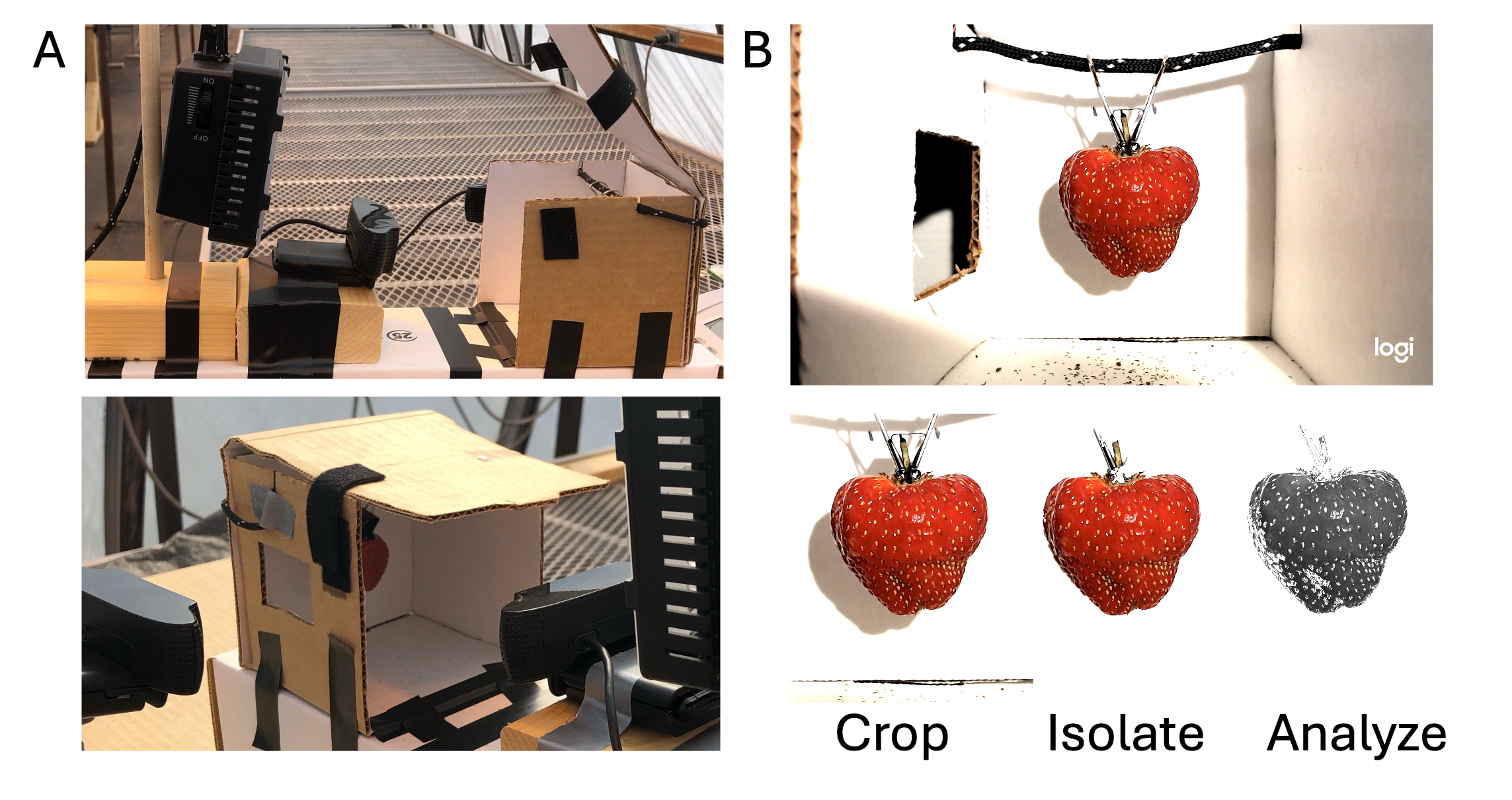}
        \caption {\small{Front and side camera setup for pictures of harvested strawberries; B) Example image from front camera in its original and then when cropped, with background removed, and in black and white for achene analysis}}
        \label{fig:CVpics}
    \end{figure}
    
\subsection{Results}
\label{sec:field_results}

    \begin{table}[b!]
        \centering        \begin{tabular}{|l|l|l|}
        \hline
        \textbf{Group 1}    & \textbf{Group 2} & \textbf{p-value} \\
        \hline
        quadcopter+Bees  & quadcopter     & 0.3397 \\
        \hline
        quadcopter+Bees  & Bees      & 0.6118 \\
        \hline
        quadcopter+Bees  & Neither   & 0.2434 \\
        \hline
        quadcopter       & Bees      & 0.9644 \\
        \hline
        quadcopter       & Neither   & 0.9929 \\
        \hline
        Bees        & Neither   & 0.8860 \\
        \hline
        \end{tabular}
        \caption{\small{Pairwise comparisons between the control groups from the field experiment}}
        \label{tab:pvalues2020F}
    \end{table}
    
To analyze the effects on the four treatment groups, each harvested strawberry was weighed on a scale with 0.01 g accuracy and photographed with Logitech webcams in a custom photo booth from two adjacent sides (see Fig.~\ref{fig:CVpics}. The pictures were then run through a script that output parameters describing the hue, symmetry, size (based on pixel-area), and level of pollination (based on the size and spacing of the achenes), similar to the work done by~\cite{ares2009development,wietzke2018insect,klatt2014bee,tafuro2022strawberry,sawada2024method}. The mean and standard deviation of these data are displayed in Table \ref{tab:2020Fdata}.
    
\begin{figure}[b!]
        \centering
        \includegraphics[width=.8\columnwidth]{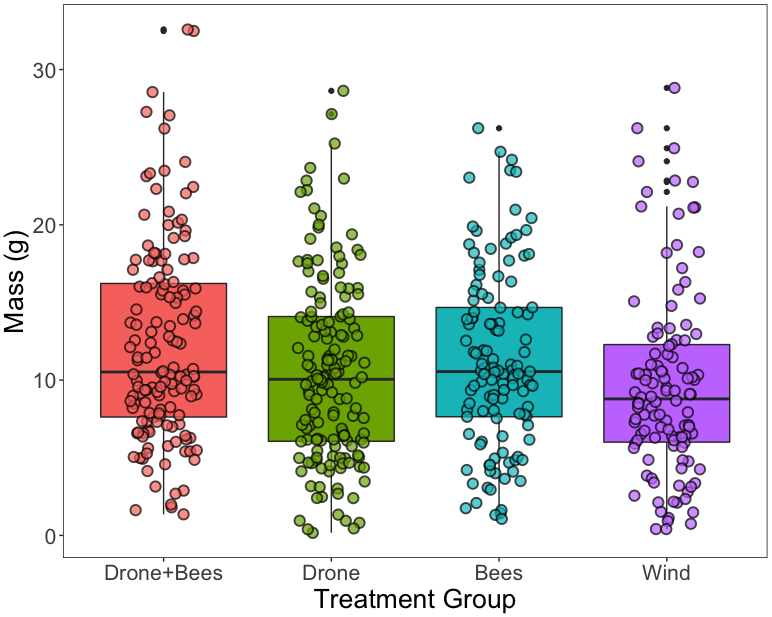}
        \caption {\small{Box and whisker plot by control group, with the mass of each berry overlaid, for the field experiment}}
        \label{fig:boxW2020F}
    \end{figure}

For symmetry, the algorithm determines the centroid of the berry and draws a line between the stem and the bottom of the strawberry. It then counts the number of pixels on either side of the major axis, and the symmetry score becomes the percent difference between the two sides. 
For size, the algorithm computes the $bw_{area}$ which is an area estimate based on the number of pixels in the berry, and then uses the picture quality (pixels/$in^2$) to convert to a measurement of surface area. 
To determine the degree of pollination, the algorithm thresholds the image and looks for boundaries matching the size and shape of an achene. The average and standard deviation of the relative distance between achenes and their size are then recorded. Fig.~\ref{fig:boxW2020F} calls out the comparison between the mass data specifically.

To determine the quality of the data, we performed several statistical analyses. After inspecting the individual summary statistics, we created a linear mixed model of the data using the treatment group as our fixed effect. We included the following random effects: plant and groups of five plants (spatial blocking to account for soil and other factors that vary along the length of the row), and temporal blocking (to account for variations in weather day-to-day). We then did pairwise comparisons on the different treatment groups using the Kenward-Roger method and 95\% confidence intervals and then calculated the power of the experiment.
The data from the field experiment had a power of 0.95. The statistical comparisons of the different treatment groups is shown in Table \ref{tab:pvalues2020F}.

\subsection{Discussion}
\label{sec:field_discussion}

The field experiment had large data sets and utilized an experimental protocol for isolating plants from pollinators that have been used in other pollinator research for strawberries~\cite{klatt2014bee}. The power of the experiment was greater than 0.9, indicating that the results can be considered meaningful. The variance of all of the random variables also indicated that their respective inclusion in the model was meaningful. 
In spite of this, the mass and other measures of the berries (Table~\ref{tab:2020Fdata}) show only minor differences in the size and quality of the berries between the four treatment groups; the quadcopter flights were not detrimental, but also did not yield the positive trends we had hoped to see at statistically significant levels. 
The data from this experiment does not indicate that either quadcopter pollination (as implemented) or visits from natural pollinators had any significant effect on the growth of strawberries compared to the growth of strawberries which were self- and/or wind-pollinated. 

The row of Malwina plants were the end row in a field with other earlier-flowering strawberry varieties that was adjacent to a fallow field (i.e. there was nothing flowering nearby); it could be that that natural pollinators favored other more flower-abundant areas instead. Without specifically monitoring visits from natural pollinators, however, it is difficult to know how much pollinator activity there was to the plants in question. 
The results could also be due to variation in strawberry dependence on natural pollinators with variety; some strawberry varieties depend more on self-pollination or wind-pollination. Malwina is a new variety and has not yet been studied for a correlation between natural pollinator visits and increases in yield, size, or appeal.

\section{Conclusion}
\label{sec:conclusion}


Our field experiment highlights some of the challenges faced by engineers in solving agricultural problems: even with significant support from subject-matter experts on pollination, strawberries, and field work, it was difficult to control for all possible confounding variables or keep animals or passing u-pickers from eating the 'fruit of our labors'.

Our computer vision approach to analyzing strawberries is not novel~\cite{ares2009development,wietzke2018insect,klatt2014bee,tafuro2022strawberry,sawada2024method}, but we are hoping to convert our analysis code into a Jupyter Notebook for open-source availability for farmers and researchers wishing to assess the level of pollination in strawberries without coding expertise. 

If we were to repeat the experiment, we would choose a variety like Jewel known to vary drastically in size with pollination levels~\cite{allen2022croppol}. We would also try to eliminate edge effects by securing a mature row of plants with neighboring rows on each side. To better isolate and monitor the control groups we would use Osmolux bags~\cite{klatt2014bee}, which would prevent wind-pollination as well as pollination by natural pollinators (leaving only self-pollination). We could also look into additional flight patterns or different types of drones.

Although the results of our field data were statistically inconclusive, the lab experiment suggested that quadcopter downdraft is capable of moving pollen-sized particles. Further study of the force required to separate pollen from the flower, particle image velocimetry analysis of the quadcopter's downdraft, the optimal number of flights/day, for example, could help us better understand whether wind can play a more significant role in strawberry pollination.

\section{Acknowledgments}
\label{sec:acknowledgments}

Thank you to Indian Creek Farms for allowing us to rent their strawberries and land for research purposes. Thank you to the Cornell Statistical Consulting Unit for helping with the statistical modeling. Thank you to Dr. Heather Grab for lending her strawberry expertise and Dr. Scott McArt for his bee expertise. Thank you also to Chad Thomas, other Cornell greenhouse staff for helping us learn to work with strawberries and attempt the experiment in the greenhouse first. Finally, thank you to Elena Suarez who worked hard on preliminary and unincorporated aspects of this project.

\bibliographystyle{IEEEtran}
\bibliography{References}

\end{document}